\documentstyle[epsf]{aipproc}
\def\gsim{\stackrel{>}{\sim}}
\def\lsim{\stackrel{<}{\sim}}
\def\beq{\begin{equation}}
\def\eeq{\end{equation}}
\def\ba{\begin{array}}
\def\ea{\end{array}}

\def\egzk{E_{GZK}}
\begin{document}
\title{Signature Studies of Cosmic Magnetic Monopoles}
\author{Stuart D. Wick$^{*,}$\footnote{Presenter at the First International
Workshop on Radio Detection of High--Energy Particles, November 16 - 18,
2000, UCLA.}, Thomas W. Kephart$^{\dagger}$, and
Thomas J. Weiler$^{\dagger}$}
\address{$^*$ Department of Physics, University of Florida,\\
Gainesville, FL 32611\\
$^{\dagger}$Department of Physics and Astronomy, Vanderbilt
University,\\
Nashville, TN 37235}
%
\maketitle
\begin{abstract}
This talk explores the possibility that the Universe may be populated
with relic magnetic monopoles. 
Observations of galactic and extragalactic magnetic
fields, lead to the conclusion that monopoles of mass $\lsim 10^{14}$
GeV are accelerated in these fields to relativistic velocities. 
The relativistic monopole signatures and features we derive are
(i) the protracted shower development, (ii) the Cherenkov signals,
(iii) the tomography of the Earth with monopoles, 
and (iv) a model for monopole airshowers above the GZK cutoff. 
\end{abstract}
\section*{Introduction}
Any symmetry breaking, after inflation, of a semisimple group 
to a subgroup leaving an unbroken $U(1)$ may produce 
an observable abundance of magnetic monopoles.  
The inferred strength and coherence size of existing
extragalactic magnetic fields suggest that any free 
monopole with a mass 
near or less than $10^{14}$ GeV would have been accelerated 
in magnetic fields to relativisitic velocities. On striking matter, 
such as the Earth's atmosphere, these relativistic monopoles will 
generate a particle cascade.  Here we investigate the 
associated shower signatures.

The free monopole flux is limited only by Parker's upper bound
$F_{\rm{P}}\sim 10^{-15}$/cm$^2$/s/sr~\cite{Parker}, which results from
requiring that monopoles not short--circuit our Galactic magnetic fields
faster than their dynamo can regenerate them.  Since the Parker bound is
several orders of magnitude above the observed highest--energy cosmic
ray flux, existing cosmic ray detectors can meaningfully search for a
monopole flux.

Because of their mass and integrity, a single monopole primary will
continuously induce air--showers, in contrast to nucleon and photon
primaries which transfer nearly all of their energy at shower
initiation. Thus, the monopole shower is readily
distinguished from non--monopole initiated showers.  We also
investigate the possibility that the hadronic cross--section 
of the monopole is sufficient to produce air--showers comparable to
that from hadronic primaries, in which case existing data would already
imply a meaningful limit on the monopole flux.  One may even speculate
that such monopoles may have been observed, as the primaries
producing the enigmatic showers above the GZK cutoff at 
$\sim~5~\times~10^{19}$~eV~\cite{PorterGoto,KW96}.

\section*{Characteristics of a Monopole Flux}

The flux of monopoles emerging from a
phase transition is determined by the Kibble mechanism \cite{Kibble}.
At the time of the phase transition, roughly one monopole or
antimonopole is produced per correlated volume, $\xi_c^3$. 
The resulting monopole number density today is
\beq
n_M \sim 10^{-19}\, (T_c/10^{11}{\rm GeV})^3 (l_H/\xi_c)^3\,{\rm
cm}^{-3},
\label{density}
\eeq
where $\xi_c$ is the phase transition correlation length, bounded from
above by the horizon size $l_H$ at the time when the system relaxes to
the true broken--symmetry vacuum. 
Although minimal $SU(5)$ breaking gives monopoles of mass $\sim 10^{17}
\;\rm{GeV}$,  there are ample theoretical possibilities 
for producing monopoles with
smaller mass while maintaining the possibility of strong interaction
cross--sections that avoid proton decay \cite{SK,Hong,Dutta,FLFK}. 
Based on the Kibble mechanism
for monopole production, bounds on the universe's curvature constrain
the monopole mass to less than $10^{13}$~GeV, while a comparison of the
Kibble flux to the Parker limit constrains the monopole mass to less
than $10^{11}$~GeV. 
The general expression for the relativistic monopole flux 
may be written \cite{KW96}
\beq
F_M = c\: n_M/4\pi \sim 2\times 10^{-16}\, \left(\frac{M}{{10^{11}{\rm
GeV}}}\right)^3\left(\frac{l_H}{\xi_c}\right)^3\,{\rm cm}^{-2} \;{\rm
sec}^{-1}\;{\rm sr}^{-1}\,.
\label{flux}
\eeq
In higher dimensional cosmologies, the Kibble flux may be altered; 
then the straightforward Parker
upper limit $F_P\leq 10^{-15}/{\rm cm}^2/{\rm sec}/{\rm sr}$ becomes the
only reliable bound on the monopole flux.  In the spirit of
generality, we take the monopole mass $M$ to be a free parameter 
and the Kibble mechanism is a rough guide to $F_M$. We require that
$F_M$ obey the Parker limit and assume that proton decay is
avoided in a way that does not restrict the parameter $M$.

\subsection*{Monopole Structure}

Monopoles are topological defects with a non-trivial internal structure;
the core of the monopole is a region of restored unified symmetry.
Monopoles are classified \cite{Kibble} by their 
topological winding, but
for the case of GUT monopoles this classification is too coarse.  In an
$SU(5)$ GUT the fundamental minimally-charged monopole is six-fold
degenerate.  For an appropriate Higgs potential there are four other
types of stable bound states formed from the fundamental
monopoles \cite{Harvey2,Tanmay2}.
This work distinguishes between those monopoles with
color--magnetic charge and those with only ordinary $U_{EM}(1)$
magnetic charge.  Thus, we adopt the nomenclature ``$q$--monopoles'' for
those monopoles with color--magnetic charge and ``$l$--monopoles'' for
those with only the ordinary magnetic charge. 

The possible confinement
of $q$--monopoles has recently been considered \cite{Goldhaber} via
the formation of $Z_{3}$ color--magnetic ``strings.'' If such a
mechanism were realized one result could be the formation of
color--singlet ``baryonic--monopoles'' in which the 
fusion of three differently
colored strings produces a baryon--like composite of $q$--monopoles. 
The internal structure of a baryonic--monopole would
approximate that of an ordinary baryon in the QCD string model, but with
$q$--monopoles in the place of the quarks.  Thus, the baryonic--monopole
structure is quite different from a single $l$--monopole and, as such,
it is shown to have a very different cross--section and cosmic ray
shower profile.

\subsection*{Monopole Acceleration}

The kinetic energy imparted to a magnetic monopole on traversing a magnetic
field along a particular path is \cite{KW96}
\beq
E_K=g \int_{\rm{path}} \vec{B} \cdot\vec{dl}\,
\sim\,g\,B\,\xi\,\sqrt{n}\,
\label{Ekin}
\eeq
where
\beq
g=e/2\alpha=3.3\times10^{-8} \;\rm{esu} \;\;
({\rm or}\; 3.3\times 10^{-8} dynes/G)
\label{charge}
\eeq
is the magnetic charge according to the Dirac quantization condition,
$B$ is the magnetic field strength, $\xi$ specifies the
field's coherence length, and $\sqrt{n}$ is
a factor to approximate the random--walk through the $n$ domains
of coherent field traversed by the path.  Galactic magnetic
fields and magnetic fields in extragalactic sheets 
and galactic clusters range from about $0.1$ to $100 \mu G$, 
while their coherence lengths
range from $10^{-4}$ to about $30\rm{Mpc}$ \cite{Kron,Ryu97}.
These fields can accelerate a monopole from rest to the energy range
$2\times 10^{20}$ to $5\times 10^{23}~\rm{eV}$. 
For extragalactic sheets the number of random-walks
can be roughly estimated to be of order $n \sim H_0^{-1}/50\,
\rm{Mpc} \sim 100$, and so $E_{\rm max} \sim 5 \times 10^{24}$~eV. 
Hence, monopoles with mass below $\sim 10^{14}$~GeV may be relativistic. 
The rest of this
talk is devoted to the novel phenomenology of relativistic monopoles. As
a prelude to calculating monopole signatures in various detectors, we
turn to a discussion of the interactions of monopoles with matter.

\section*{Relativistic Monopole Energy Loss}

Both $l$--monopoles and baryonic--monopoles are conserved
in each interaction because of their topological stability.
However, as conjectured above, their different internal structures
will lead to differing shower profiles and signatures.
Because of space limitations, in this talk we only consider
the electromagnetic interactions of $l$--monopoles and the
hadronic interactions of baryonic--monopoles.

The shower profile of baryonic--monopoles is based upon a 
model \cite{monotross} where the hadronic cross--section grows
after impact and the net energy transfer is enough to stop
the monopole very quickly.  Since this mechanism is model
dependent, we consider the baryonic--monopole signatures less
reliable.  Further discussion of the 
baryonic--monopole is postponed until the final section.

Our most reliable signatures are for $l$--monopoles (which
are referred to as ``monopoles'' for the remainder of this
talk) and are based upon well understood electromagnetic 
processes.  At large distances and high velocities, 
a monopole mimics the
electromagnetic interaction of a heavy ion of charge 
$Z\sim 1/2\alpha\simeq 68$.
We view the monopole as a classical source of radiation, while
treating the matter--radiation interaction quantum mechanically. In
this way, the large electromagnetic coupling of the monopole is isolated
in the classical field, and the matter--radiation interaction can be
calculated perturbatively.

\begin{figure}[b]
\centerline{\hbox{
\vspace{6.5 in}
\epsfxsize=255pt \epsfbox{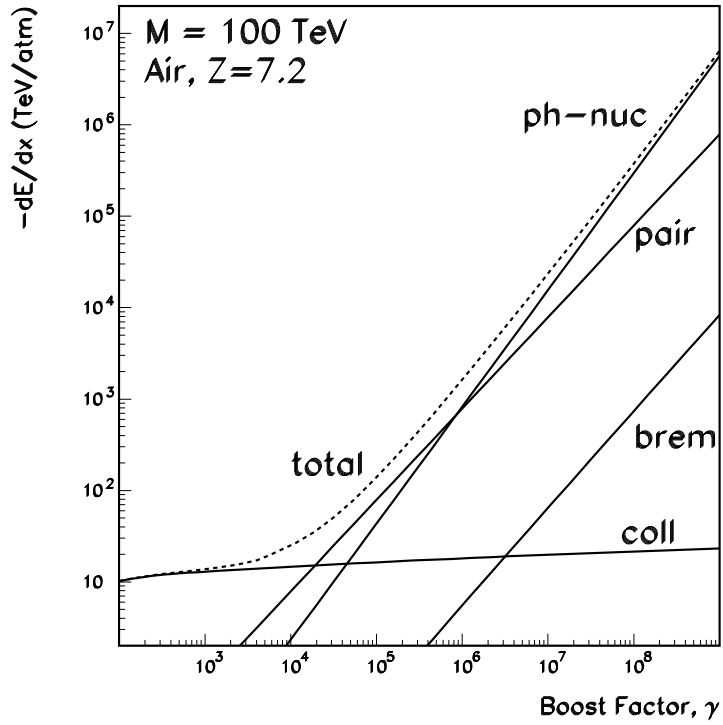}
}}
\caption{The electromagnetic energy loss from collisional, bremsstrahlung,
electron--pair production, and the photonuclear interaction 
of a $100$ TeV relativistic monopole in air.  Collisional, 
pair production, and the photonuclear interaction are roughly
independent of the monopole mass whereas bremsstrahlung is
$\propto M^{-1}.$  The units of energy loss are given in 
TeV per atmosphere.}
\label{fig:eloss}
\end{figure}

%
\subsection*{Electromagnetic Interactions}

We consider here the energy loss of a monopole resulting from four
electromagnetic processes: collisions (ionization of atoms), $e^+ e^-$
pair production, bremsstrahlung, and the photonuclear interaction. 
All of these processes involve the scattering of a virtual photon,
emitted by an incident monopole, off of a target particle. 

The monopole--matter electromagnetic interaction for monopole boosts 
$\gamma < 100$ is well reported in the literature \cite{Giac1,Ahlen1}.  
Previous works include atomic excitations and ionization losses, 
including the density suppression effect.  These
processes are collectively referred to as ``collisional'' energy losses
and are $\propto \ln\gamma.$ 
The pair production ($MN\rightarrow MNe^+e^-$) 
and bremsstrahlung ($MN\rightarrow MN\gamma$) energy losses are 
$\propto \gamma, $ where $M,N,e$ and $\gamma$ represent a monopole,
nucleus, electron, and photon respectively. 
The photonuclear ($MN\rightarrow MNX,$ where $X$
are hadrons) energy loss \cite{Reno} is roughly $\propto \gamma^{1.28}.$ 
For large $\gamma,$ the pair production and photonuclear
interactions dominate while bremsstrahlung is suppressed 
by the large monopole mass as $M^{-1}.$  (By comparison,
the bremsstrahlung of a muon is of similar strength to other 
radiative energy loss processes.)

Here we only have space to collect the 
electromagnetic energy loss processes
together and plot them, in fig.~(\ref{fig:eloss}),
for $M=100$ TeV monopoles (see \cite{monotross} for more details).

\section*{Monopole Electromagnetic Signatures}

Signature events for monopoles are discussed with a specific emphasis on
1) the general shower development, 2) the direct Cherenkov signal, 
3) the coherent radio--Cherenkov signal, and 4) the tomography of the
Earth's interior.  Monopoles
will be highly penetrating primaries, interacting via the
electromagnetic force and all the while maintaining their structural
integrity.  On average, there will be a quasi-steady cloud of secondary
particles traveling along with the monopole.  Thus, 
we will call this type of shower ``monopole--induced.''

\begin{figure}[b]
\centerline{\hbox{
\vspace{6.5 in}
\epsfxsize=255pt \epsfbox{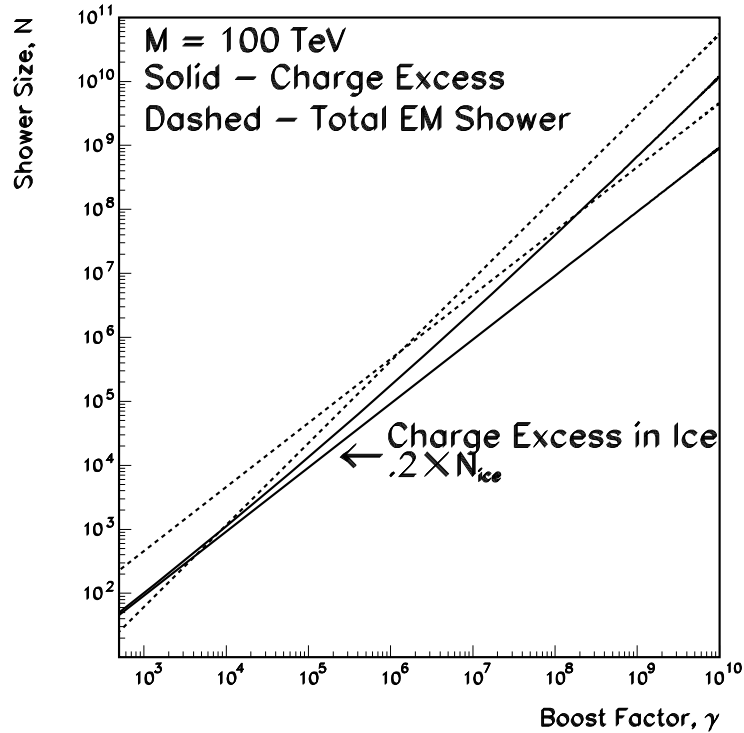}
}}
\caption{The monopole-induced quasi-steady shower 
size in ice for a monopole of mass $100$ TeV.  The shower
size is the total number of electron, positrons, and
photons.  The dashed line $\propto \gamma$ is for pair
production alone and the dashed line $\propto \gamma^{1.28}$ 
is for the photonuclear interaction alone.
The solid lines show the electric charge excess 
(roughly 20\% of the shower size) for pair production alone
($\propto \gamma$)
and pair production plus photonuclear ($\propto \gamma^{1.28}$).}
\label{fig:shower}
\end{figure}

Given a fast monopole passing through matter, the various
electromagnetic processes can inject energetic photons, electrons, 
positrons, and hadrons into the absorbing medium.
If the energy of these injected secondary particles is sufficient
(roughly greater than $E_{c}\sim 100$ MeV), they
may initiate a particle cascade.  
In terms of the inelasticity $\eta~\equiv~\Delta E/E,$
the condition for electromagnetic shower development 
is $\eta \gsim E_{c}/E_{0} \simeq
10^{-12}\left( E_{0}/10^{20}\rm{eV}\right)^{-1}.$  
Lower inelasticity events will contribute directly to
ionization without intermediate particle production.  
The inelasticity per interaction and the subsequent
shower development is best understood for pair production.
Detailed calculations \cite{monotross} show 
that for $\gamma \gsim 10^{4}$ all of the monopole energy
lost via pair production goes into the electromagnetic
shower.

The contribution of the photonuclear process to the 
{\it{electromagnetic}}
shower is indirect.  The photonuclear interaction 
injects high energy hadrons
into the monopole--induced shower.
A subshower initiated by a high energy 
hadron will produce $\pi^{0}$'s as secondaries, which each decay 
to 2 $\gamma$'s.  If these $\gamma$'s have $E > E_{c},$ 
they may initiate an electromagnetic shower.  
So, only a fraction the energy lost via the 
photonuclear interaction contributes to the
electromagnetic shower in the end.

Given the arguments above, it is reasonable to 
assume that pair production 
alone provides a lower bound to the electromagnetic
shower size and that the pair production plus photonuclear
interaction provides an upper bound.  
We plot the pair production and photonuclear processes 
separately (dashed lines) in fig.~(\ref{fig:shower}). 

The electromagnetic shower sweeps a net charge excess
from the medium into the shower of roughly 20\% the shower size.
For the charge excess we are again justified in using 
pair production alone as a 
lower bound and pair production plus photonuclear as an
upper bound.  This is reflected in 
fig.~(\ref{fig:shower}) by plotting pair production alone (the 
solid curve $\propto \gamma$) and by plotting pair production plus 
the photonuclear interaction (the solid curve $\propto \gamma^{1.28}$).

The lateral profile is approximately 
uniform out to a lateral cutoff given by the Moli\`{e}re radius
\beq
R_{\rm{M}}=7.4\,\frac{\rm{g}}{\rm{cm}^{2}}
\;(\frac{\xi_{e}}{35 {\rm
g/cm^2}})\,(\frac{100 \rm{MeV}}{E_{c}}),
\label{eq:moliere}
\eeq
where $\xi_{e}$ is the electron radiation length.
As defined, the Moli\`{e}re radius is independent of the incident
primary energy, being determined only by the spread of low energy
particles resulting from multiple Coulomb scattering.  Within a distance
$R_{\rm{M}}$ of the monopole path will be $\sim$ 90\% of the shower
particles \cite{Partdata}.

\subsection*{Monopole Cherenkov Signatures}

When a charge travels through a medium with index of refraction $n,$ at
a velocity $\beta> 1/n,$ Cherenkov radiation is emitted.  
The total power emitted in Cherenkov radiation per unit
frequency $\nu$ and per unit length $l$ by a charge $Ze$ is given by the
Frank-Tamm formula
\beq
\frac{d^{2}W}{d\nu \,dl} = \pi\alpha Z^{2} \nu\left[ 1 -
\frac{1}{\beta^{2}n^{2}} \right].
\eeq
The maximal emission of the Cherenkov light occurs at an
angle $\theta_{\rm{max}}=\arccos(1/n\beta)$ where $\theta$ is
measured from the radiating particle's direction.
The interaction of a magnetic charge
with bulk matter requires the replacement of factors of $\epsilon$ with
the Maxwell dual factors $\mu$. But $\mu$ and $\epsilon$ are related 
by the index of refraction. The replacement in the electric
charged--particle interaction formulae (for $Z=1$) 
adequate for magnetic monopoles is
$\alpha \rightarrow n^2/4\alpha$, and
leads to an enhancement factor of 4700 for monopoles
interacting in vacuum and 8300 for monopole interactions in water. 
However, in matter a relativisitic monopole is accompanied by an
extensive
cloud of charged particles it continually produces, so the difference
in monopole electromagnetic interactions caused by the index of
refraction factor is totally obscured.

The monopole-induced shower also contributes to the Cherenkov
signal.  In particular, the electric charge excess (of roughly
20\% the shower size as shown in fig.~(\ref{fig:shower})) 
will emit coherent Cherenkov for radio
wavelengths, $\lambda >> R_{\rm{M}}.$  For coherent radio--Cherenkov
the $Z^{2}$ factor could be large, with $Z^{2} \lsim 10^{18},$
while the shower size is expected to remain roughly constant
as the monopole traverses a large--scale ($\sim \rm{km}^{3}$) 
detector.  Thus, a monopole
signature event is clearly distinct from that of a neutrino
event in the RICE array or similar large--scale detectors.
The non--detection of a monopole event after one year of
observation in a $\sim \rm{km}^{3}$ detector can, conservatively,
set a flux limit of
\beq
F_M \; \lsim \; 10^{-18} \;{\rm cm}^{-2} \;{\rm
sec}^{-1}\;{\rm sr}^{-1}\,
\label{RICElimit}
\eeq
which is significantly below the Parker limit. 

\subsection*{Earth Tomography with Relativistic Monopoles}
Direct knowledge about the composition and density of the Earth's
interior is lacking. Analysis of the seismic data is currently the best
source of information about the Earth's internal properties
\cite{LayGeo}. However, another potential probe would be the study of
highly penetrating particles which could pass through the Earth's
interior and interact differently depending upon the composition and
density of material traversed.   Thus, it may be possible to
directly measure the density profile of the
Earth's interior \cite{foot8}. Over a range of masses,
$M \sim 10^{4 \pm 1}$ TeV, and initial kinetic energies, 
monopoles can pass through the
Earth's interior and emerge with relativistic velocities and,
therefore, function as such probe.  See \cite{monotross} for more details.

\section*{Baryonic--Monopole Air Showers}

The natural acceleration of monopoles to energies above the GZK cutoff
at $\egzk\sim 5\times 10^{19}$~eV, and the allowed abundance of a
monopole flux at the observed super--GZK event rate motivates us to ask
whether monopoles may contribute to the super--GZK events. As a proof of
principle, we have studied a simple model of a baryonic--monopole
interaction in air which produces a shower similar to that arising from
a hadronic primary. To mimic a hadron--initiated shower the 
baryonic--monopole must
transfer nearly all of its energy to the shower over roughly a
hadronic interaction length, $\lambda_{0} 
\sim 80 \;\rm{g}\;\rm{cm}^{-2}.$ 
The large inertia of a massive monopole makes this impossible
if the cross--section is typically  strong, $\sim 100$~mb
\cite{Minertia}.  The cross--section we seek needs to be much larger.

We model our arguments on those of \cite{Goldhaber} where three
$q-$monopoles are confined by $Z_{3}$ strings of color--magnetic flux to
form a color--singlet baryonic--monopole.  We further assume that 1) the
cross--section for the interaction of the baryonic--monopole with a
target nucleus is geometric; in its unstretched state (before hitting the
atmosphere) the monopole's cross--section is roughly hadronic,
$\sigma_{0} \sim \Lambda^{-2}$ (where $\Lambda\equiv\Lambda_{QCD}$); 2)
each interaction between the baryonic--monopole 
and an air nucleus transfers an
$O(1)$ fraction of the exchanged energy into stretching the
chromomagnetic strings; 3) the chromomagnetic strings
can only be broken with the formation of a monopole--antimonopole pair,
a process which is highly suppressed and therefore ignored; other
possible relaxation processes of the stretched string are assumed to be
negligible; 4) the energy transfer per interaction is
soft, $\Delta E/E~\equiv~\eta~\sim~\Lambda/M$. 

The
color--magnetic strings have a string tension $\mu \simeq \Lambda^{2}$.
Therefore, when $O(1)$ of the energy transfer ($\gamma \Lambda$)
stretches the color--magnetic strings (assumption 2), the length $l \sim
\Lambda^{-1}$ increases by $\delta l = dE/\mu$, so that the fractional
increase in length is  $\delta l/l=\gamma$. Consequently, the
geometrical cross--section grows 
$\propto \gamma\Lambda^{-2}$ after each
interaction. The energy loss for baryonic-monopoles can then be 
approximated as
\beq
\frac{dE}{dx}(x)\;\simeq\;
-\frac{\gamma\,\Lambda}
{\lambda(x)}\;\simeq\; - \gamma\,\Lambda
\,n_{N}\,\sigma(x),
\label{eq:hadstop2}
\eeq
where the strong cross--section $\sigma (x)$ is
explicitly a function of column depth $x$ and $n_{N}$ is the
number density of target nucleons. 
From assumption (4) we infer that
the total number of monopole-nucleus interactions required to
transfer most of the incoming kinetic energy is roughly
$\eta^{-1}.$
From the above discussion, the geometrical cross--section
after $n$ interactions is
\beq
\sigma_{n}\sim\frac{1+\sum_{i=1}^{n}
\gamma_{i}}{\Lambda^{2}}
=\frac{1+n\gamma}{\Lambda^2}\,,
\eeq
where we have approximated $\gamma_n\sim (1-\eta)^n\,\gamma\sim \gamma$.
The mean-free-path $\lambda\equiv 1/ \sigma n_{N}$
after the $n$-th interaction is therefore
\beq
\lambda_{n}\sim \frac{\Lambda^{2}}{n_{N}\gamma n},
\hspace{1cm}n\geq 1\,,
\label{eq:mfp}
\eeq
and the total distance traveled between the first interaction and
the $(\eta^{-1})^{\rm th}$ interaction is
\beq
\Delta X \sim \sum_{n=1}^{\eta^{-1}}\lambda_{n}\, 
\sim \frac{\Lambda^{2}}{n_{N}\,\gamma}
\ln\left(\frac{M}{\Lambda}\right)\,<<\lambda_{0} 
\eeq
for $\eta^{-1}\gg 1.$
Thus the stretchable chromomagnetic strings of the baryonic--monopole
provide an example of a very massive monopole which nevertheless
transfers O(1) of its kinetic energy to an air shower over a
very short distance.  In conclusion, the baryonic--monopole's 
air--shower signature roughly mimics that of a hadronic primary.

\section*{Acknowledgments.}

\noindent This work was supported in part by the U.S. Department of
Energy grants no. DE-FG05-86ER40272 (SDW),
DE-FG05-85ER40226 (TWK \& TJW), and the Vanderbilt University Research
Council.


\begin{references}
\bibitem{foot8}This idea has been exploited in neutrino physics as
neutrinos are sufficiently weakly interacting to pass through the earth
largely unimpeded for neutrino energies $\lsim 10^{15}\;\rm{eV}.$
\end{references}
\end{document}